# Electrodeposition of amorphous molybdenum sulfo-selenide as a low-cost catalyst

Lee Kendall, Giovanni Zangari, Stephen McDonnell

Abstract:

Overall, the $MoS_x$ system has shown greater catalytic activity over the traditional $MoS_2$ based systems due to the absence of discrete basal plane, and differing structural arrangements. This increases the overall catalytic site density alongside adequate electronic conductivity from short-range atomic arrangements that allow for use in electrochemical processes. Here we translate prior efforts to improve electrocatalysis via Se incorporation within the crystalline system to the polymeric system, as several of the active sites in the a-$MoS_x$ have similar motifs and bonding environments as those found in $MoS_2$. We use a single-electrolyte electrodeposition synthesis technique in order to provide a scalable, low-cost material. We demonstrate the influence of the electrolyte conditions on the films physical, chemical, electronic, and catalytic properties as a function of selenium content through a comprehensive study via Raman spectroscopy, X-ray photoelectron spectroscopy (XPS), UV-Vis spectroscopy, and electrochemical methods. Together, these results highlight selenium's role in the Se-$MoS_x$ system in promoting the catalytic efficacy of these catalysts.

Intro:

$MoS_{2x}Se_{2(1-x)}$ has recently been discussed as a potential candidate for the use as a catalytically active material for HER.[1-3] This is partly due to several experimental studies that demonstrated the increased catalytic performance of $MoS_{2x}Se_{2(1-x)}$ alloys, where it was shown that the HER performance of crystalline $MoS_{2x}Se_{2(1-x)}$ was significantly superior to those of 2H-$MoS_2$, or of $MoSe_2$.[2] Indeed, the addition of a different chalcogen atoms produced new active sites in the basal plane of $MoS_2$ via defect generation.[1] Additionally, DFT studies have examined the electronic structure and hydrogen adsorption energy for $MoS_{2x}Se_{2(1-x)}$ and determined that the band gap for the alloyed material was lower than that of the semiconducting 2H-$MoS_2$ and 2H-$MoSe_2$ by reducing the conduction band minimum (CBM).[2, 3] The mechanisms for improving the electrocatalysis of the alloy are two-fold. First, the lowered bandgap boosts the HER performance by increasing the conductivity of the material.[3] Second, it has been reported that facilitation of hydrogen adsorption on transition metal dichalcogenides is strongly dependent to the conduction band minimum, rather than the traditional d-band center.[4, 5] Overall, these improvements to the alloy over the binary constituents lead to significantly more thermoneutral values for the sulfur and selenium sites.[2] This work aims to apply these results of increased HER performance conducted within the crystalline system to that of a polymeric $MoS_x$ system. The polymeric $MoS_x$ system has shown greater catalytic activity over the traditional $MoS_2$ based systems, due to the absence of discrete basal plane and differing structural arrangement that increases the overall catalytic site density alongside with adequate electronic conductivity from short-range atomic arrangements, that allow for use in electrochemical processes.[6-11]

**Methods:**

Electrolyte Preparation

The synthesis of Se-$MoS_x$ catalysts were prepared with a chemical bath containing $Mo^{6+}$and $S^{2-}$, with the addition of $Se^{2-}$. The electrolyte was prepared from sodium molybdate dihydrate ($Na_2MoO_4 \cdot 2H_2O$), sodium sulfide hydrate ($Na_2S \cdot H_2O$), and selenium dioxide ($SeO_2$). $SeO_2$ was chosen for its abundance and solubility within water and sulfur rich electrolytes. We hypothesize the sulfidation/selenization of molybdate ions to the chalcogen-based ions follow a four-step reaction described by the equation:

$$MoO_4^{2-} + HS^-/HSe^- + 4H^+ <-> (MoS_{4-x}Se_x)^{2-} + 4H_2O$$

The pH of the electrolyte was acidified from its initial pH of roughly 13 down to a pH of 8. This was done to help facilitate the formation of the final reactants over the intermediates. The $Na_2MoO_4$ and $Na_2S$ were added prior to any titration whereas the $SeO_2$ was added to the electrolyte only after the pH had reached ~9. This was done as it was observed that $SSe_2$ precipitates would preferentially form at higher pHs when $SeO_2$ was added before titration. To circumvent this, the $SeO_2$ was added later in titration. The electrolyte is clear prior to titration. Upon adding small amounts of $SeO_2$, there is no immediate difference. However, at high concentrations of $SeO_2$, the precipitation of $SSe_2$ is unavoidable as the increased Se concentration eventually overwhelms the buffer. Three sample sets where synthesized using 50 nM, 200 mM, and 400 mM of $Na_2S$ respectively. In each sample set a range of $SeO_2$ concentrations where evaluated included 0 mM, 5 mM, 15 mM, 25 mM, 50 mM, and 100 mM $SeO_2$.

To investigate the electrolyte prior to electrodeposition, UV-Vis spectroscopy was carried out immediately following synthesis as well as after two days as can be seen in S1. The peak at 225 nm is attributed to the $(MoO_4)^{2-}$ anion and matches literature well.[12] The peaks measure at 463 nm, 312 nm, and 207 nm have been attributed to the $(MoS_4)^{2-}$ ion, the peak at 395 nm to the $(MoS_3O)^{2-}$, and the peak at 287 nm to the $(MoS_2O_2)^{2-}$, matching well with literature.[12-14] However, note that for all oxysulfide absorption peaks, there is a significant overlap between some and/or all the various chemical states. As such, the peaks have been labelled with the primary absorption species, but it is difficult to say if only one species is present at a time. Regarding any $(MoSe_4)^{2-}$ species, there is sparce literature to compare to, with the most recent reports being from 1975.[12, 15, 16] The spectra of $(MoSe_4)^{2-}$ has two primary peaks at 550 nm and at 358 nm and are not readily apparent in the spectra. There is no literature for UV-Vis of oxyselenides, but their color has been described in literature as being orange while pure $(MoSe_4)^{2-}$ is described as red-violet. Additionally, one would expect a blue shift towards lower wavelengths as Se is added. As can be seen in Figure S1, there is little evidence of $(MoSe_4)^{2-}$ in the UV-Vis spectra. This refutes the hypothesis that the reaction will spontaneously occur as it did for $(MoS_4)^{2-}$, however, this isn't completely surprising. The open circuit potential of the electrolyte is ~ -0.21 $V_{SHE}$, which is in the range for $(MoS_4)^{2-}$ to form but not complete $(MoSe_4)^{2-}$ anions[17-21]. However, the electrolyte still looks visually different depending on the concentration of added Se as shown in Figure S2a. To rule out any influence from the Se precursor, $SeO_2$, UV-Vis spectra was collected which can be seen in Figure S2b. As can be seen, there is some overlap at short wavelengths, but no overlap between the peaks that are indicative of the other species present.

**Electrodeposition of Se-MoS$_x$**

The films were electrodeposited on (0001)-textured polycrystalline ruthenium (80 nm) sputtered onto a Si (100) wafer or an indium tin oxide (ITO) coated polyethylene terephthalate (PET) substrate. The films were deposited under potentiostatic conditions at -1.55 V vs $Hg/Hg_2SO_4$ for 300 seconds on Ru and 10 seconds on ITO/PET unless noted otherwise. The samples deposited on Ru were used in XPS, Raman, and electrochemical measurements while the samples deposited on ITO/PET were used for UV-Vis measurements. Raman spectroscopy was collected using a Renishaw Raman instrument with a 514 nm laser excitation, 1800 grating, 50x objective magnification. Normalization of the data sets was achieved by finding the spectral minimum between 700-750 $cm^{-1}$ and the spectral maximum between 300-450 $cm^{-1}$ where the normalized value at a given pixel was the pixel value minus the minimum divided by the maximum minus the minimum. X-ray photoelectron spectroscopy (XPS) spectra were collected from a Scienta Omicron R3000 using monochromatic Al Kα emission with an excitation energy of 1486.7 eV and a pass energy of 50 eV. All features were fit with Shirley backgrounds Voigt functions. UV-Vis spectroscopy was conducted with an Agilent Cary 5000 UV-Vis-NIR Spectrophotometer in a dual beam transmission configuration. Background subtraction was done using the built-in program.

**Results**

*Chemical Analysis*

XPS was conducted on every sample to analyze the chemical states of Mo, S, and Se species. A representative spectrum can be seen in Figure 1 with the rest of the deconvoluted data shown in Figure S3, Figure S4, and Figure S5. The XPS of the Mo 3d spectra were deconvoluted into binding energies of four doublets ($3d_{5/2}$ and $3d_{3/2}$) of $Mo^{4+}$, $Mo^{5+}$, Mo*, and $Mo^{4+}$ with singlets of the various S 2s states and Se 3s as shown in Figure 2.

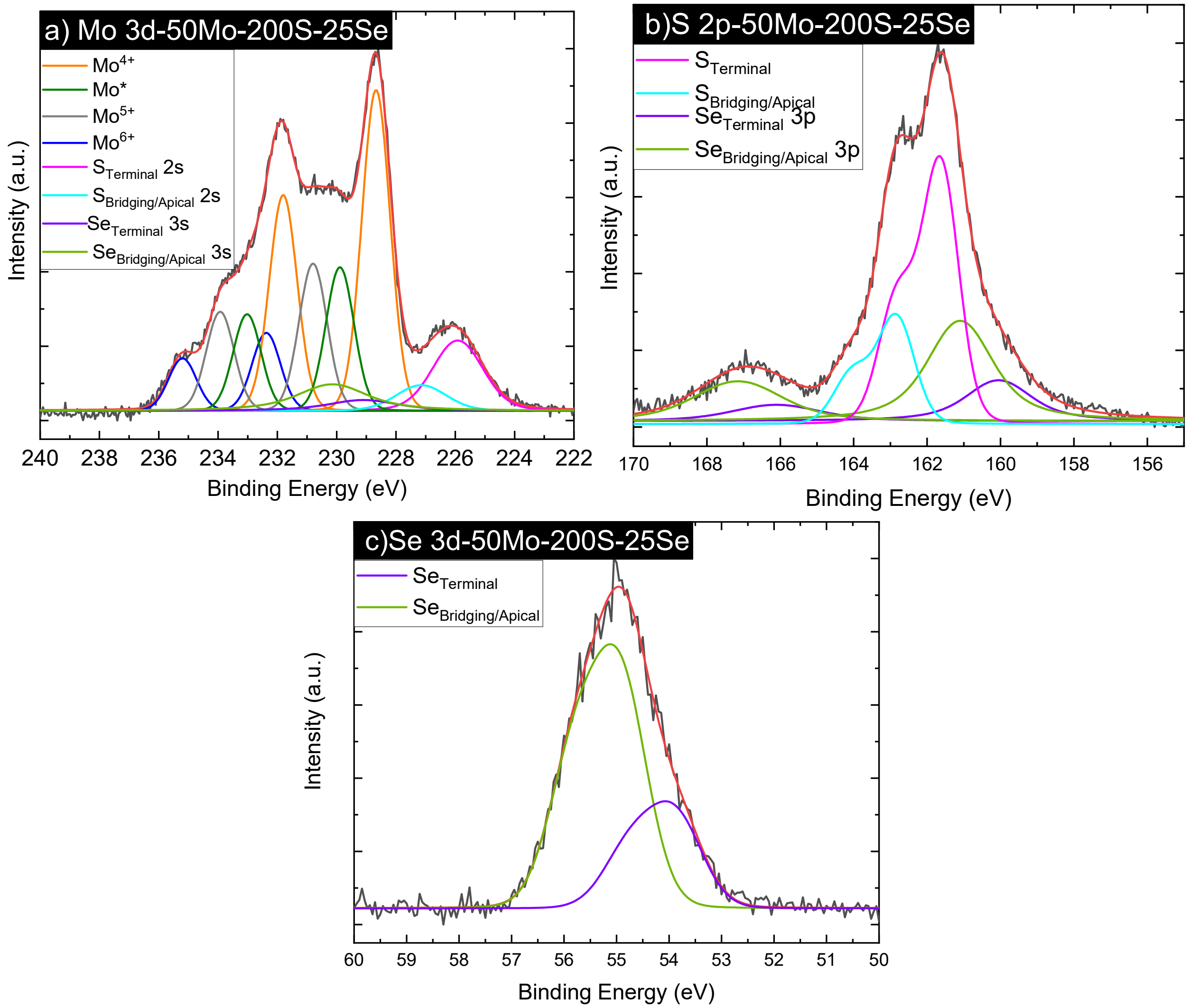

Figure 1: XPS of (a) the Mo 3d spectra, (b) the S 2p spectra, and (c) the Se 3d spectra of a representative sample

As the oxidation number of Mo increases, the binding energies at which the respective doublets are found agree with previous literature.[7, 22-26] $Mo^{4+}$ is attributed to the chalcogen saturated Mo center within the polymeric chain of $MoS_x(Se_y)$, $Mo^{5+}$ has been attributed to oxidation of the molybdenum-sulfide bonds, primarily terminal bonds as found in prior literature where Mo* has previously been attributed to an intermediate state of incomplete oxidation between $Mo^{4+}$ and $Mo^{5+}$.[23] $Mo^{6+}$ has been attributed to the Mo center of $MoO_3$. The Mo 3d peaks were deconvoluted with a binding energy difference of 3.13 eV between the $3d_{5/2}$ and $3d_{3/2}$ and with a fixed intensity ratio of 3:2. The sulfur 2s is also present within the Mo 3d spectra and is fit by restricting parameters to be consistent with the S 2p spectra. The deconvolution of the S 2p spectra consists of three primary features: (i) terminal $S_2^{2-}$ ligands (ii) bridging $S_2^{2-}$ and apical $S^{2-}$ ligands, and (iii) the overlapping Se 3p spectra as can be seen in Figure 1b. The sulfur spectrum was deconvoluted with two doublets, with the constraints of an energy difference of 1.18 eV between the $2p_{3/2}$ and $2p_{1/2}$ and with an intensity ratio of 2:1. The overlapping Se 3p spectra was fitted using fitting parameters obtained from electrodeposited Se and the peak position and intensity of the Se 3p was referenced against the primary peak of Se 3d. The Se 3d spectra was fitted with two doublets that represent the $Se^{2-}$ and $Se_2^{2-}$ ligand.[27-30] The Se spectra was deconvoluted with two doublets, with the constraints of an energy difference of 0.86 eV and an intensity ratio of 0.735 between the $3d_{3/2}$ and the $3d_{5/2}$. The S 2p and Se 3d spectra are represented here as two summed together doublets for ease of visualization.

Using the XPS spectra the overall ratio of Se to S in the film can be obtained for the various samples, as can be seen in Figure 2a. These reported values are the total value of molybdenum bonded selenium and sulfur bonds. The addition of more selenium precursors in the electrolyte generally lead to an increase in incorporated selenium found within the film. However, at a given selenium concentration, a colloidal suspension of $SeS_2$ formed and can be seen in Figure 2b and c which matches literature well.[31, 32] The increased presence of sulfur precursor delays this formation due to NaOH forming a basic buffer and preventing the precipitation of $SeS_2$ which is caused in the presence of a weak acid.[33] This formation of a suspension lead to a decrease in selenium found within the film as excess selenium is removed from the active electrolyte and therefore it likely does not participate in the deposition reaction.

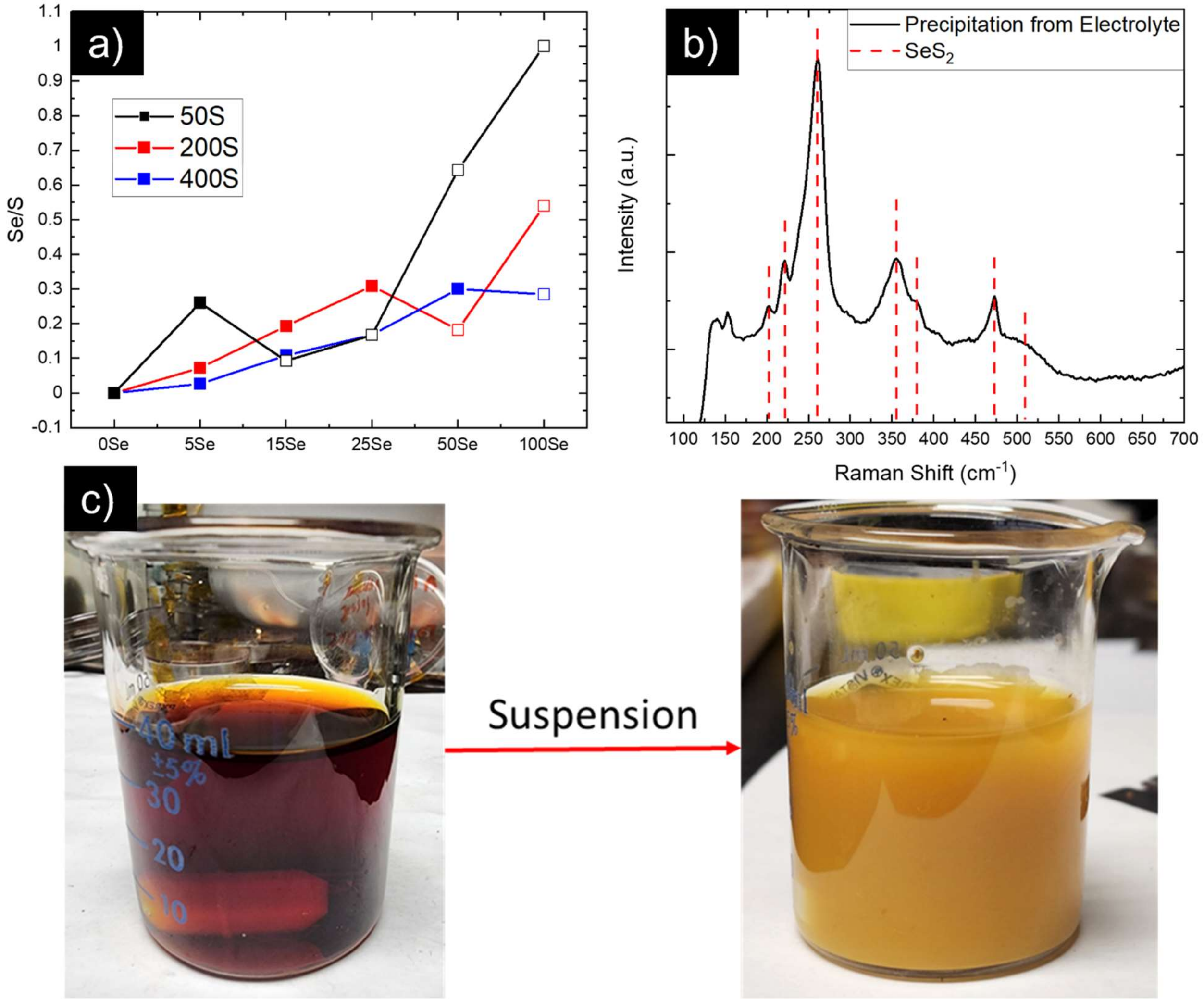

Figure 2: Se/S ratio obtained via XPS for the various films as a function of sulfur precursor concentration and as a function of selenium precursor concentration. The open boxes represent samples where the electrolyte had formed a suspension. b) Raman spectra of the filtered and rinsed precipitate. C) The formation of the suspension upon adding excess $SeO_2$

To confirm that alloying occurred within the Se-$MoS_x$ film, the binding energy of the chalcogen bonded $Mo^{4+}$ was plotted as a function of Se/S as seen in Figure 3a. It is expected that $Mo^{4+}$ shifts towards a lower binding energy as the selenium concentration increases due to the less electronegative selenium atoms.[34, 35] This further suggests that the electronic structure is being modified by varying the Se/S ratio, something that will be explored later. Additionally, it can be seen in Figure 3b that the Se is preferentially filling the apical/bridging bond-moiety, consistent with the reduction in those vibration modes found in the Raman spectra.

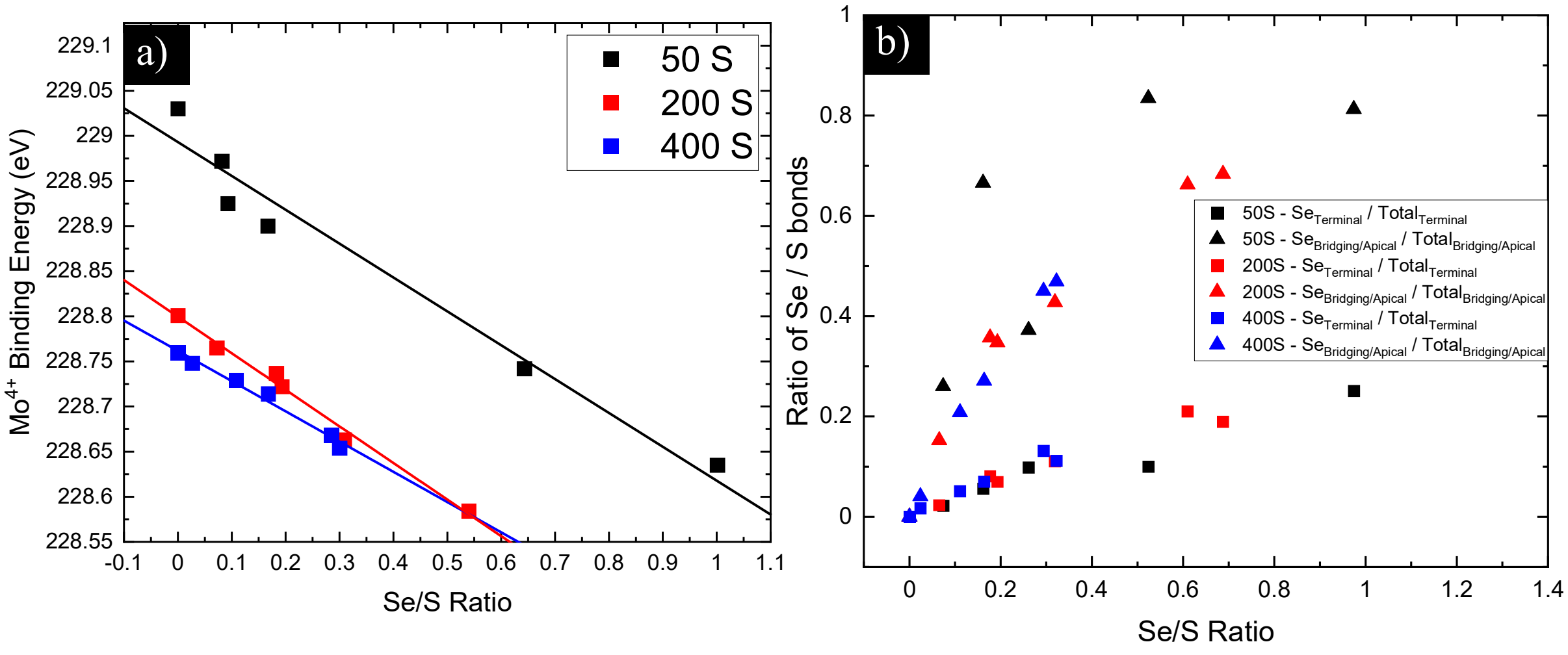

Figure 3: a) Binding energy of $Mo^{4+}$ as a function of Se/S ratio and b) Ratio of Se/S bonds as a function of Se/S. The triangles represent the ratio of $Se_{Bridging/Apical}$ / $Total_{Bridging/Apical}$ whereas the boxes represent the ratio of $Se_{Terminal}$ / $Total_{Terminal}$. The different colors represent the different sample sets.

*Structural Analysis*

Raman spectroscopy was used to investigate the structure of the various Se-$MoS_x$ electrocatalysts and can be seen in Figure 4. The Raman signatures of the Se-$MoS_x$ material shows very similar Raman signatures to those reported for $[Mo_3S_{13}](NH_4)_2$ clusters.[6, 36] The molybdenum sulfide ν(Mo-S) were found at 283-385 cm$^{-1}$ and the ν(Mo-$S_{apical}$) vibrations were observed in the 450-475 cm$^{-1}$ region where the ν(Mo-S-Mo) vibrations were observed around 425 cm$^{-1}$. The ν(S-S)$_{terminal}$ and ν(S-S)$_{bridging}$ were observed around 520 cm$^{-1}$ and 550 cm$^{-1}$, respectively. Additionally, there is a peak around 195 cm$^{-1}$ that has been recorded but not identified in previous literature. Further compounding is the lack of reports that include wavenumbers down in this range.[6, 9, 37] The peak at 255 cm$^{-1}$ could either be attributed to amorphous selenium or to a ν(Mo-Se) bond. Both may contribute to the presence of this peak, but as amorphous selenium was not observed in any of the XPS data (the $3d_{5/2}$ would be at ~55.5 eV), it is unlikely. However, it has been reported that amorphous selenium build-up is caused by the irradiation of the laser on the sample[214]. To determine any impact this has on the resulting spectra, we studied the Raman spectra as a function of laser irradiation time and found the peak evolving over time, suggesting the accumulation of amorphous selenium as seen in Figure 4e. This phenomenon was observed within the amorphous $MoSe_x$ system.[30]

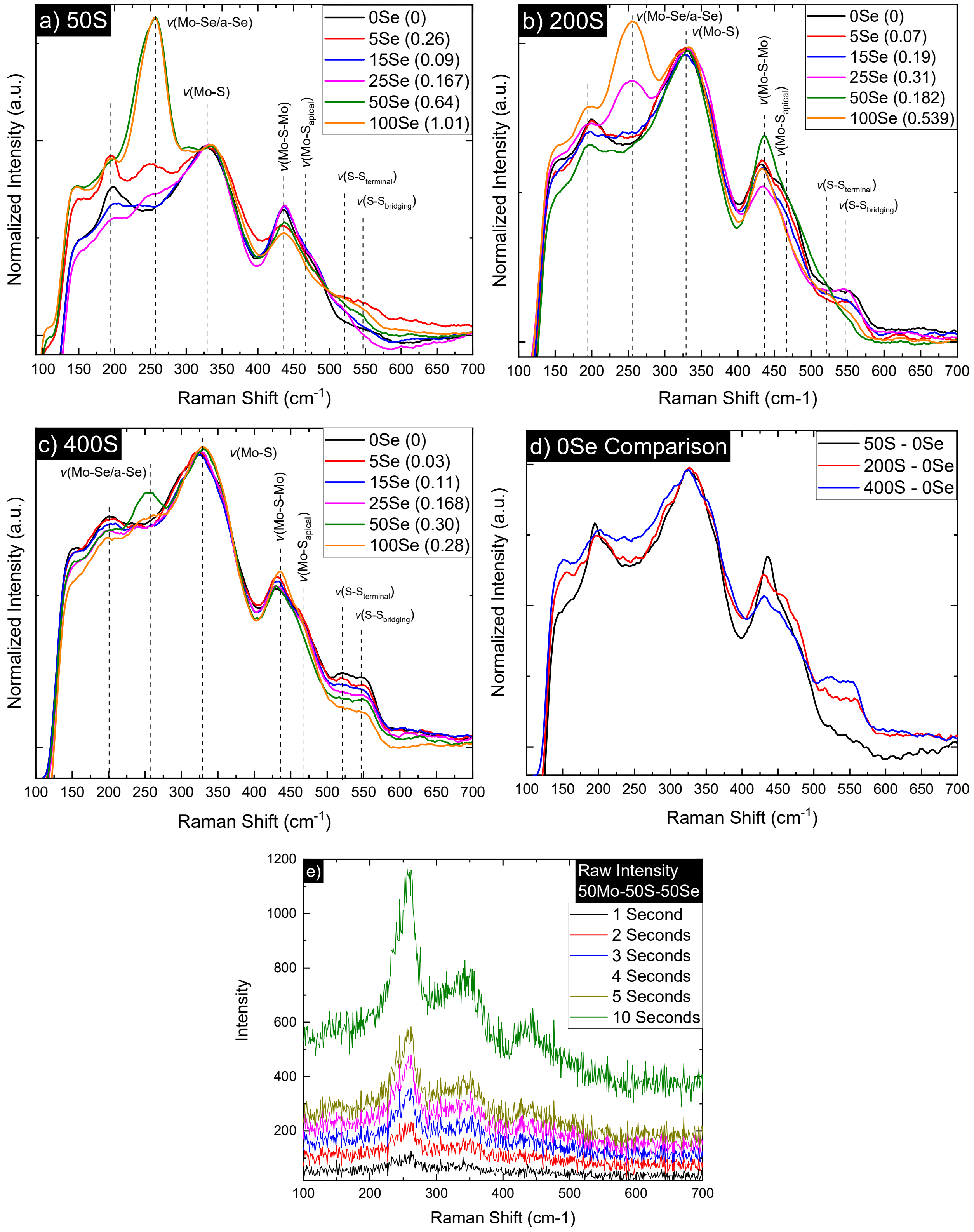

Figure 4: Raman of the various Se-$MoS_x$ electrocatalysts as a function of the Se precursor (the Se/S ratio is given in parenthesis) with a) 50S, b) 200S, and c) 400S sample sets. d) is a comparison between the various films prior to any addition of $SeO_2$ and f) is a time series of the 50Mo-50S-50Se sample where the scans were summed for the given amount of time to demonstrate any changes as a function of laser irradiation

Other changes are found within the (Mo-S-Mo), ($S-S_{terminal}$), and ($S-S_{bridging}$) bonds. When comparing these bonds between the different samples sets, it is clear that the starting, non-selenium incorporated samples, are distinctly different from each other, and therefore each sample set has a distinctly different structure which can be seen in Figure 4d. Notably, the 50 S sample has the highest concentration of (Mo-S-Mo) modes relative to (Mo-S) and it decreases in intensity as the sulfur concentration increases. The ($S-S_{terminal}$) and ($S-S_{bridging}$) vibration modes are the lowest for the 50 S sample and increase in intensity as the sulfur concentration increases to 200 S and 400 S. This can be tied back to the initial electrolyte in Figure S1 as the initial 400S electrolyte has stronger absorbance peaks for the $(MoS_4)^{2-}$ anion compared to 50S. This demonstrates that the low sulfur samples have unsaturated bonds, something that has been reported in literature.[9, 23] Within the 50 S sample set, there are subtle changes within the (Mo-S-Mo) and ($Mo-S_{apical}$) vibration modes. The main changes can be seen in the ($S-S_{terminal}$) and ($S-S_{bridging}$) vibration modes where the intensity and broadening generally increases with increased selenium content. The main spectral changes within the 200 S sample set are a suppression of the ($S-S_{terminal}$) and ($S-S_{bridging}$) bonds. A similar trend is found within the 400 S sample set. Within literature, electrodeposited $MoSe_x$ does not exhibit vibration modes at these wavenumbers, potentially explaining the reduction in these bonds. The correlation between incorporated Se, these bonds, and the catalytic properties will be discussed later.

*Optical Characterization*

To investigate possible changes in electronic structure, UV-Vis spectroscopy was conducted and can be seen in Figure 5. There is a primary absorption peak between 360 nm and 385 nm, and a secondary peak from the ITO,[38] is located around 320 nm. Additionally, interference patterns between the electrodeposited films and the ITO substrate can be seen within the low absorption regions. The primary absorption peak has been reported in literature previously and has been attributed to the direct transition from the deep valence band of S 2p to the conduction band of Mo 3d.[39-41] As the sulfur concentration increases, the primary absorption peak shifts towards higher wavelength, suggesting that the increased saturation of sulfur bonds found within the Raman spectrum leads to this shift. A similar shift has been reported in literature where the undercoordinated molecule is shifted to the left with respect to a fully saturated molecule.[13] Performing a Tauc analysis, as shown in Figure 5b,d,f, the optical direct band gap transition can be calculated.

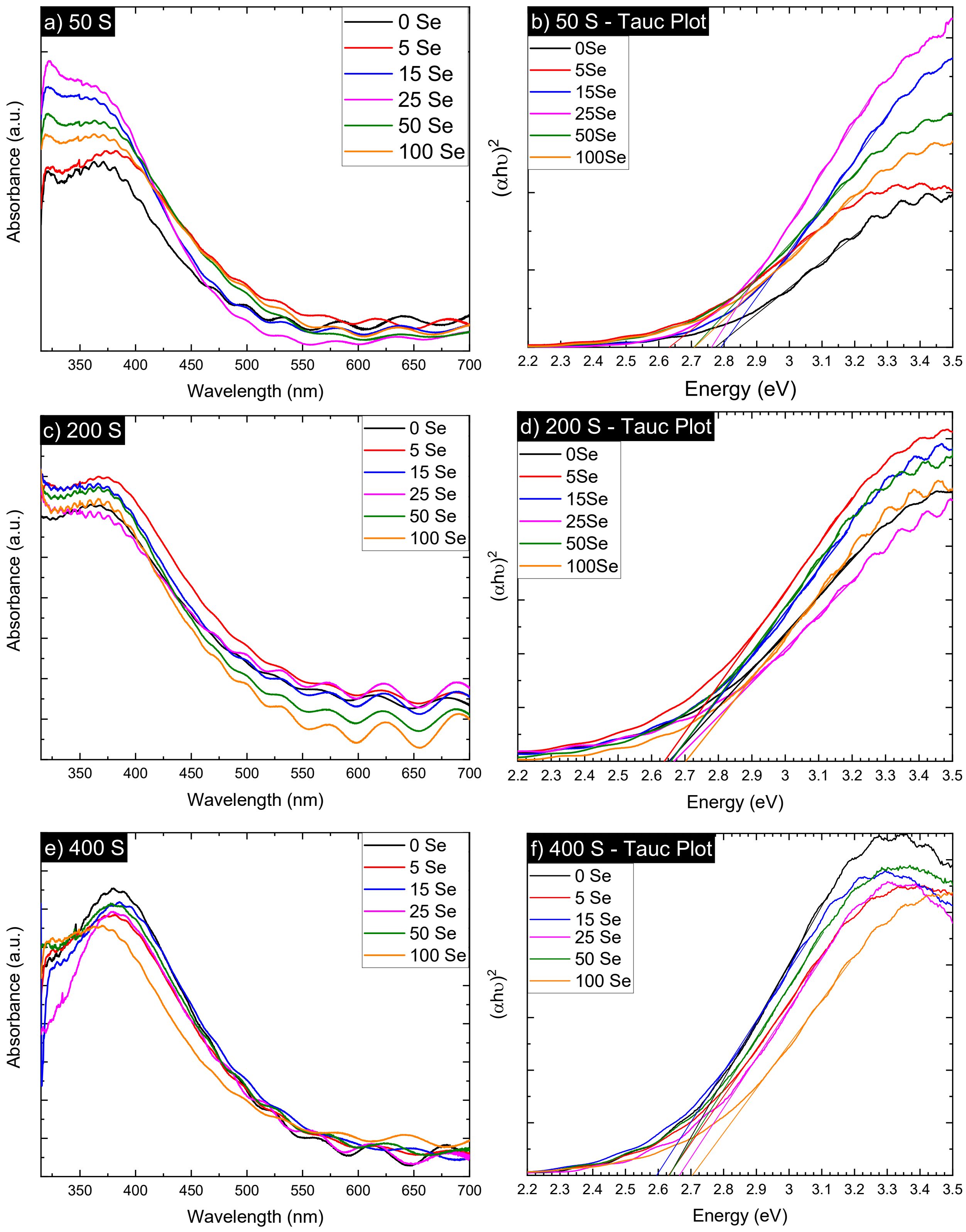

Figure 5: UV-Vis spectrum for the various sample sets, with a) 50 S, c) 200 S, and e) 400 S with the corresponding Tauc plots to the right, with b) 50S, d) 200S, and c) 400S

These calculations are plotted as a function of incorporated Se as shown below in Figure 6 and are what is expected for a-$MoS_x$ based materials.[39] What is seen is a slight decrease in the direct band gap value for the low Se/S ratios, following which the band gap quickly rises after a certain point. This trend, along with the lack of a large overall change, was expected based on previous literature and density of states calculations.[42-44]

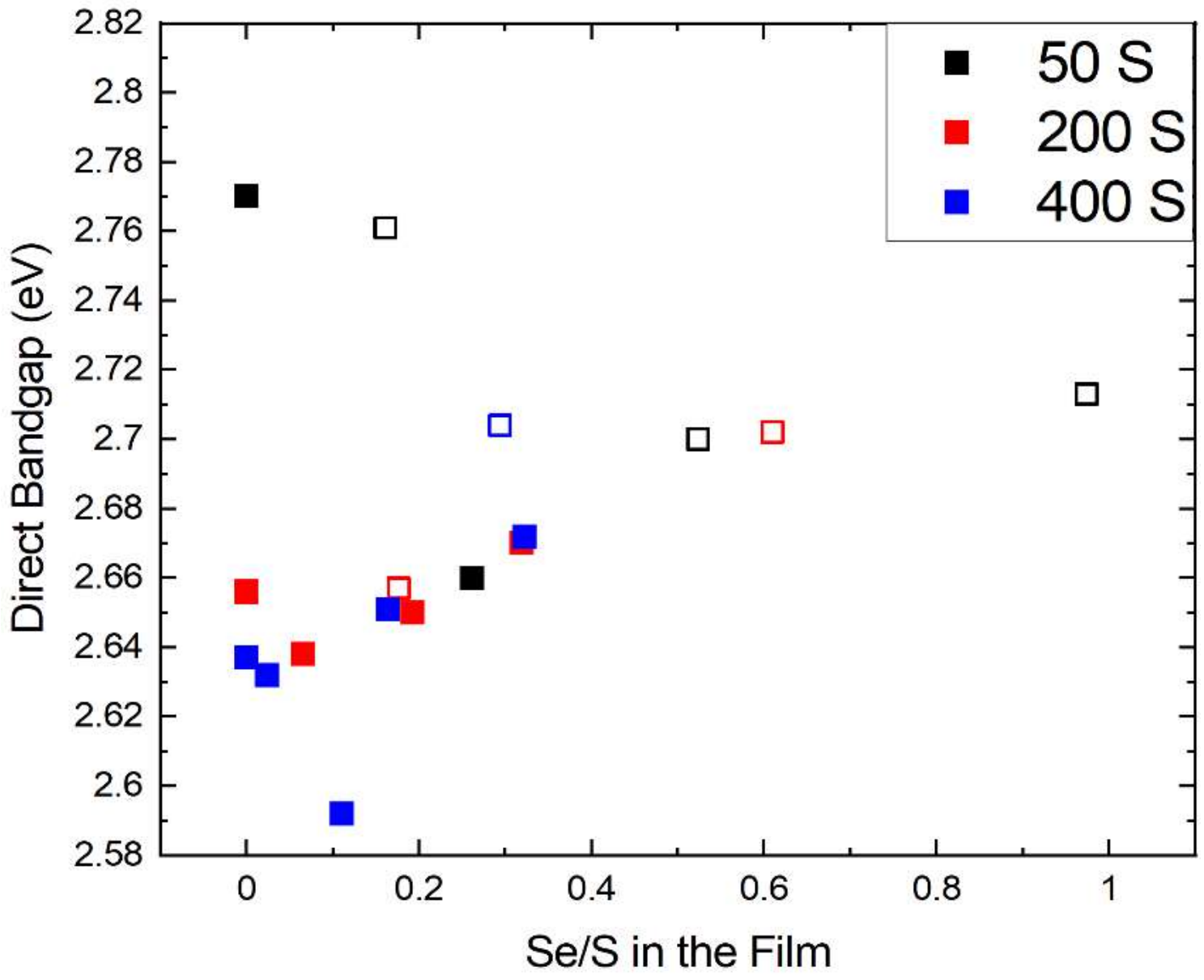

Figure 6: Graphical representation of the direct bandgap calculated via Tauc plots as a function of Se/S in the film, note: open boxes demonstrate when a suspension was present.

*Electrocatalytic properties extraction*

To examine the impact that the selenium incorporation into $MoS_x$ has on the catalytic performance for HER, polarization experiments were conducted and are shown in Figure 7. To compare the various catalysts, the overpotential required to reach 10 mA/cm$^2$ ($\eta_{10}$) and the Tafel slope are recorded and can be seen in Table 1. As can be seen, the sample that exhibited the lowest overpotential was the 400 S – 15 Se sample at only 153 mV to reach 10 mA/cm$^2$, which is competitive with the best a-$MoS_x$ style materials, many of which require a 3D support to reach this activity. Additionally, the sample is quite stable as shown in Figure 7e, improving over the standard $MoS_x$ catalyst from previous work.[23, 45] The overpotentials as well as the associated Tafel slopes are summarized in Table 1. Regarding the Tafel slope, there are two possible combinations for the overall mechanism, the Volmer-Tafel and the Volmer-Heyrovsky. If the adsorption of the hydrogen was the limiting step, a Tafel slope of 120 mV/dec would be anticipated. If the Tafel or Heyrovsky mechanisms are the limiting step, a Tafel slope of 30 or 40 mV/dec are expected, respectively. As such, the majority of the samples follow the Volmer-Heyrovsky mechanism, consistent with previous reports of a-$MoS_x$-based materials.[6, 9] Of note, when the precipitation reaction occurs in 50 S – 15 Se, 200 S – 50 Se, and 400 S – 100 Se, the Tafel slope increases significantly. The catalytic properties of the Se-$MoS_x$ initially increase with increasing concentrations of Se, however, decrease significantly once higher concentrations of Se are incorporated as seen in Figure 7d

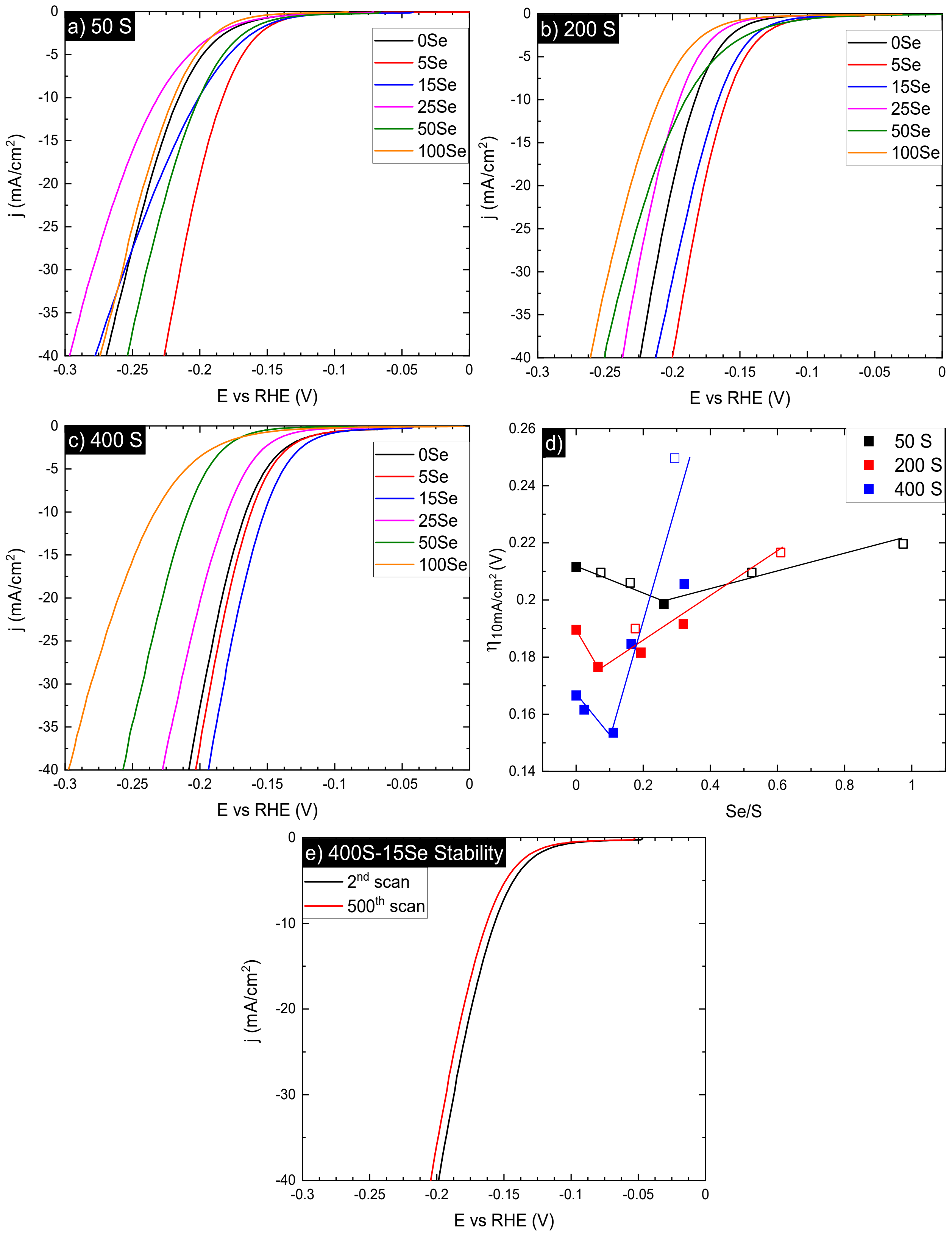

Figure 7: Polarization curves for the various samples sets, with a) 50 S, b) 200 S, and c) 400 S. In d) the overpotential is plotted as a function of Se/S ratio. The lines are there to guide the reader and are not linearly fit lines. e) stability of the best performing catalyst, 400S-15Se

Table 1: Overpotentials and Tafel slopes of the catalysts

| **$\eta_{10}$ (mV)**<br>Tafel slope (mV/dec) | 50mM $Na_2S$ | 200mM $Na_2S$ | 400mM $Na_2S$ |
|---|---|---|---|
| 0mM $SeO_2$ | **212 mV**<br>56 mV/dec | **190 mV**<br>43 mV/dec | **166 mV**<br>51 mV/dec |
| 5mM $SeO_2$ | **198 mV**<br>43 mV/dec | **177 mV**<br>45 mV/dec | **162 mV**<br>47 mV/dec |
| 15mM $SeO_2$ | **209 mV**<br>71 mV/dec | **182 mV**<br>43 mV/dec | **153 mV**<br>46 mV/dec |
| 25mM $SeO_2$ | **206 mV**<br>68 mV/dec | **192 mV**<br>44 mV/dec | **185 mV**<br>44 mV/dec |
| 50mM $SeO_2$ | **210 mV**<br>52 mV/dec | **190 mV**<br>73 mV/dec | **206 mV**<br>42 mV/dec |
| 100mM $SeO_2$ | **220 mV**<br>47 mV/dec | **217 mV**<br>52 mV/dec | **250 mV**<br>72 mV/dec |
| | 71 mV/dec | 43 mV/dec | 46 mV/dec |
| 25mM $SeO_2$ | **206 mV**<br>68 mV/dec | **192 mV**<br>44 mV/dec | **185 mV**<br>44 mV/dec |
| 50mM $SeO_2$ | **210 mV**<br>52 mV/dec | **190 mV**<br>73 mV/dec | **206 mV**<br>42 mV/dec |
| 100mM $SeO_2$ | **220 mV**<br>47 mV/dec | **217 mV**<br>52 mV/dec | **250 mV**<br>72 mV/dec |

When comparing the initial catalytic properties of the samples prior to introducing selenium, 50 S had the poorest activity with 400 S having the greatest. This is likely due to the increased saturation of the ($S\text{-}S_{bridging}$) and ($S\text{-}S_{terminal}$) bonds found in the high S samples. It has been reported in literature that these bonds, and especially ($S\text{-}S_{bridging}$), are the main sites for HER catalytic activity and therefore the more saturated those bonds are, the higher the catalytic activity.[46] This lack of sulfur bonding can be seen in both the XPS and Raman, where in XPS there is subdued $Mo^{4+}$ and in the Raman spectra there is a lack of ($S\text{-}S_{bridging}$) and ($S\text{-}S_{terminal}$) modes. At high Se/S concentrations, the ($S\text{-}S_{bridging}$) and ($S\text{-}S_{terminal}$) have been significantly replaced by selenium, leading to worse catalytic performance. This is expected as electrodeposited a-$MoSe_x$ has worse catalytic performance than similarly synthesized a-$MoS_x$ so as it becomes more $MoSe_x$ like, performance is expected to suffer.[30] Additionally, as the ($S\text{-}S_{bridging}$) are the most catalytically active, the preferential replacement of the bridging modes decreases the catalytic efficacy of those samples.

However, this does not completely explain the increase in performance for those alloys at low Se/S ratios. It is known that by lowering the band gap, the catalytic activity should increase as conductivity is critical in electro-catalysis.[3] To examine this, the $\eta_{10}$ has been plotted against the direct optical bandgap as seen in Figure 8a. It demonstrates that the 200 S and 400 S have a relationship between the overpotential required and direct optical bandgap. This is expected as the lowered bandgap would increase the conductivity of the sample, allowing for increased electron transfer. However, the 50 S sample set does not seem to be significantly impacted by the changing band gap. We hypothesize that this occurs because of the lack of saturation in the bonding environment, overriding any positive benefit towards increasing the catalytic activity that decreasing the bandgap would yield. This highlights that the catalytic sites need to be sufficiently saturated before small changes in the bandgap will demonstrate any significant improvement to the catalytic efficacy of the catalyst.

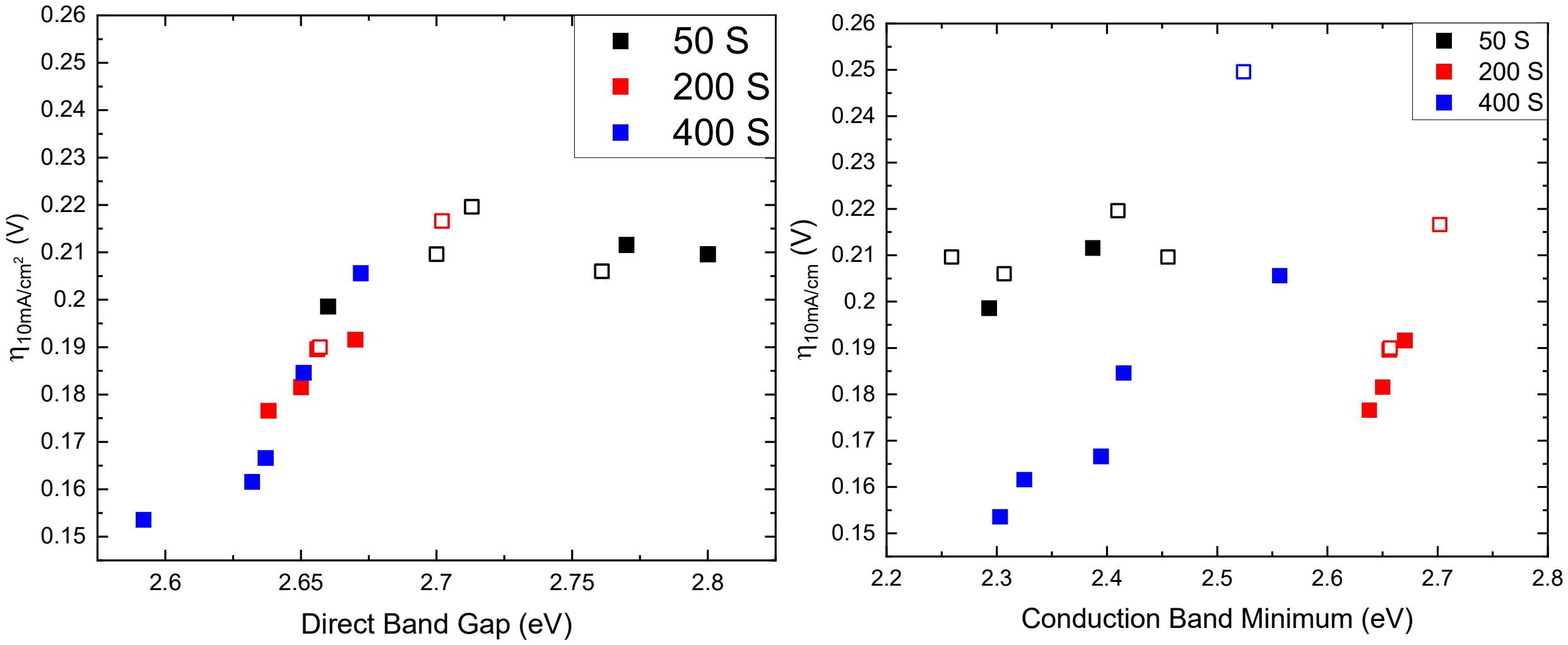

Figure 8: a) The overpotential of the various catalysts as a function of their direct optical bandgap and b) the overpotential as a function of the conduction band minimum obtained via XPS and the Tauc analysis

It has been reported that within transition metal dichalcogenides the reduction of the conduction band minimum (CBM) is attributed to the increased HER catalytic activity due to the increase of overlapping states with the states of adsorbed hydrogen. To examine this, the $\eta_{10}$ has been plotted against the conduction band minimum that was calculated from the valence band maximum obtained via XPS, seen in Figure S6, and the optical bandgap from the Tauc analysis. A similar trend from the direct bandgap figure holds, where the reduction of the CBM impacts the catalytic activity of the 200 S and 400 S samples, however it does not seem to significantly alter the 50 S sample set. This further demonstrates that the catalytic efficiency cannot be solely increased by reducing the bandgap or CBM, but that there must be sufficient catalytic sites to take advantage of the synergistic impact.

## Conclusions

In summary, we have synthesized a series of Se-$MoS_x$ films across a variety of selenium and sulfur precursor concentrations utilizing a potentiostatic electrodeposition technique. Our results demonstrate that we are able to control the chemical composition, structure, electronic properties, and catalytic efficacy of the resulting film by tailoring our starting electrolyte composition. This study has demonstrated that the saturation of the ($S\text{-}S_{bridging}$) and ($S\text{-}S_{terminal}$) bonds are critical to the HER catalytic properties and that by substituting them with low amounts of Se, the catalytic properties can be enhanced. However, when in high concentration, the films catalytic properties suffer. Additionally, this study investigated the impact of Se incorporation has on the bandgap and CBM of the resulting films and ties it to catalytic properties, supporting prior research that the films with the smallest bandgap and lowest CBM exhibited the greatest catalytic performance if the active sites are present. Overall, an active catalyst that only requires an overpotential of 153 mV with a low Tafel slope of 46 mV/dec is reported and takes advantage of synergistic properties of alloying a-$MoS_x$ with Se.

Acknowledgements:

The authors acknowledge the Nanomaterials characterization facility (NMCF) for access to characterization tools and also Shaui Li and Prof. Mona Zebarjadi for training and access to the Agilent Cary 5000 UV-Vis-NIR Spectrophotometer.

Supporting Information

Electrodeposition of amorphous molybdenum sulfo-selenide as a low-cost catalyst

Lee Kendall, Giovanni Zangari, Stephen McDonnell

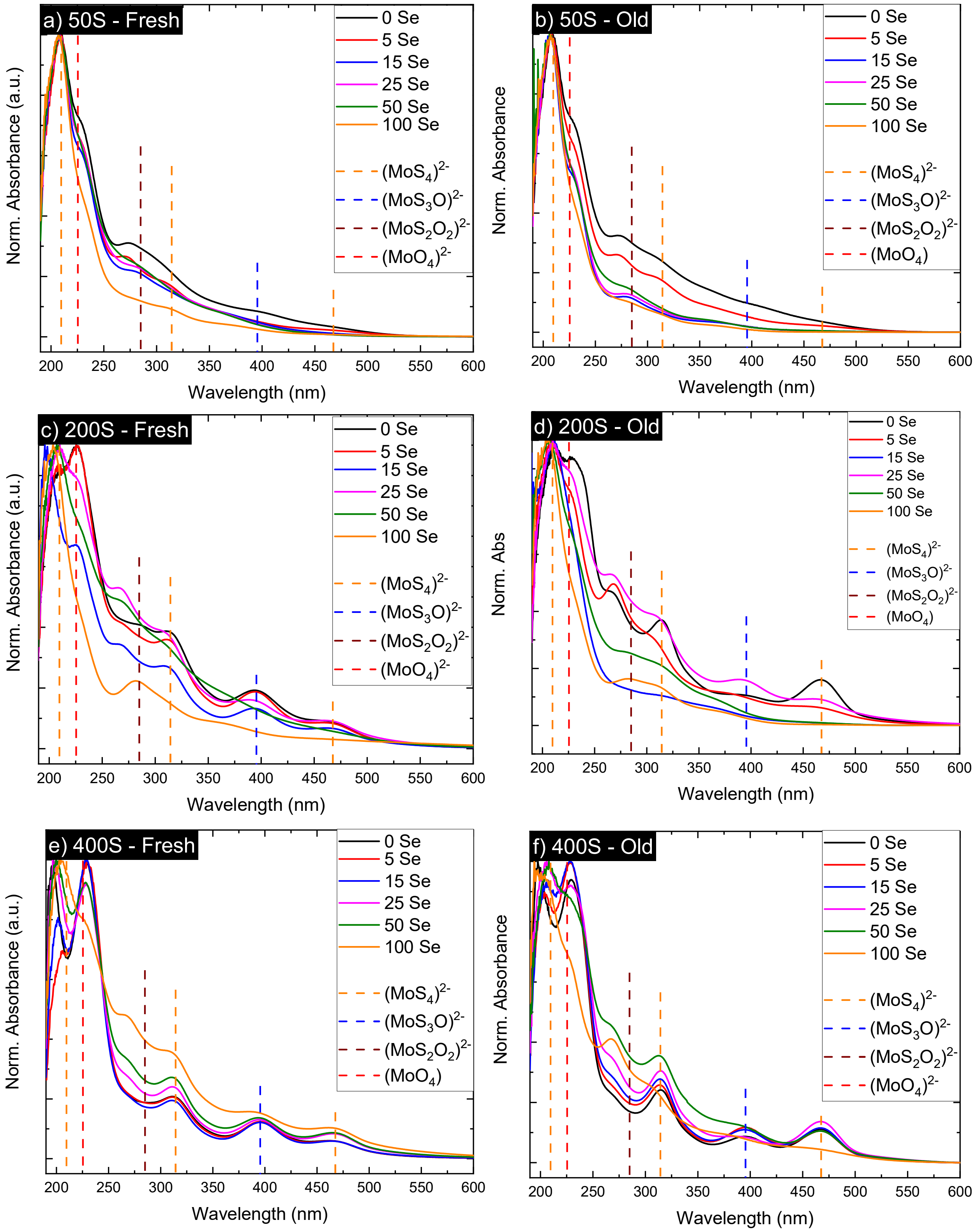

Figure S1: UV-Vis spectra of Se-$MoS_x$ electrolytes: a) sample set with 50 mM $Na_2S$ immediately after synthesis, b) sample set with 50 mM $Na_2S$ after two days, c) sample set with 200 mM $Na_2S$ immediately after synthesis, d) sample set with 200 mM $Na_2S$ after two days, e) sample set with 400 mM $Na_2S$ immediately after synthesis, f) sample set with 400 mM $Na_2S$ after two days

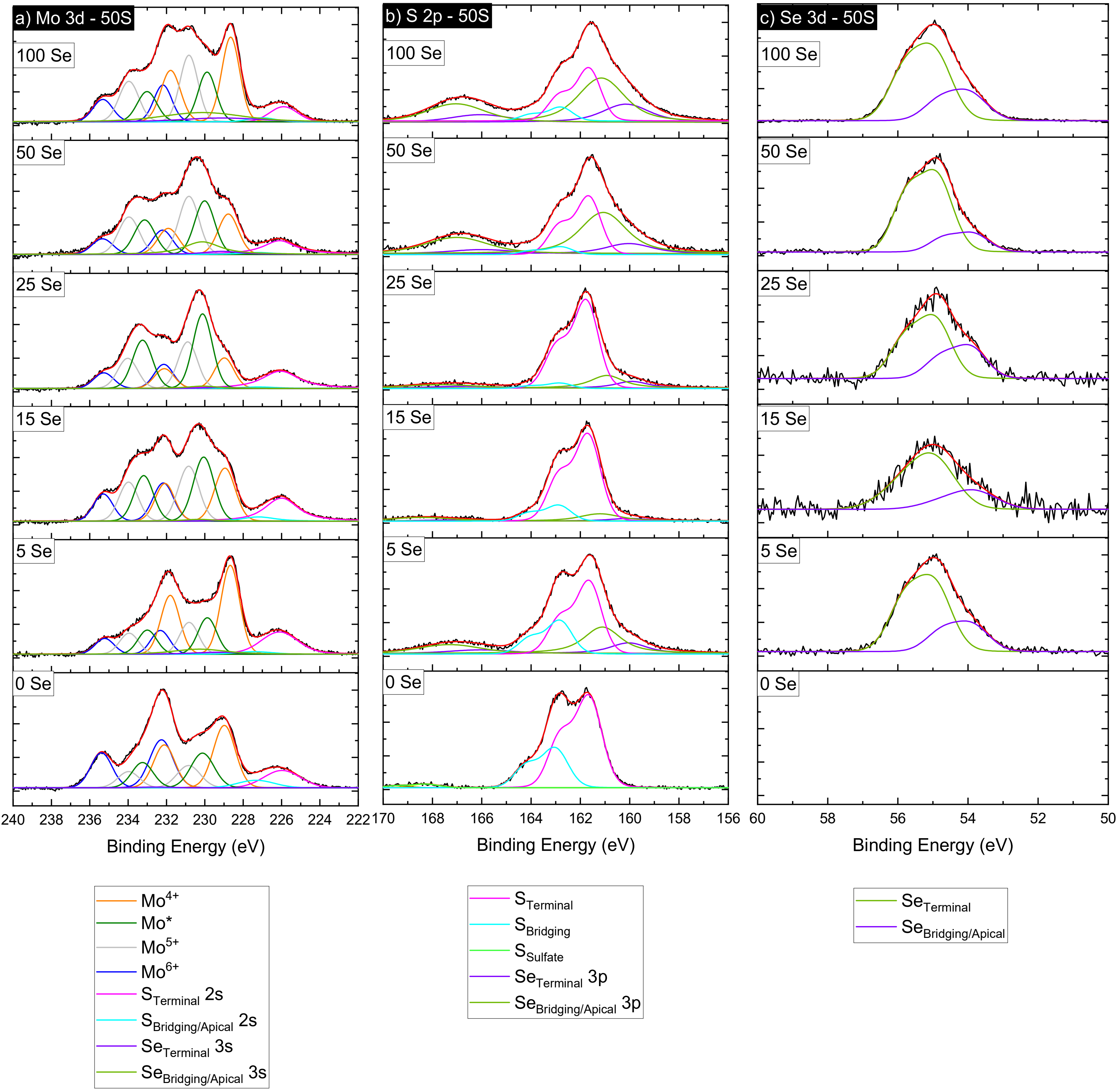

FigureS2: XPS of all samples within the 50S sample set, with a) Mo3d, b) S 2p, and c) Se 3d

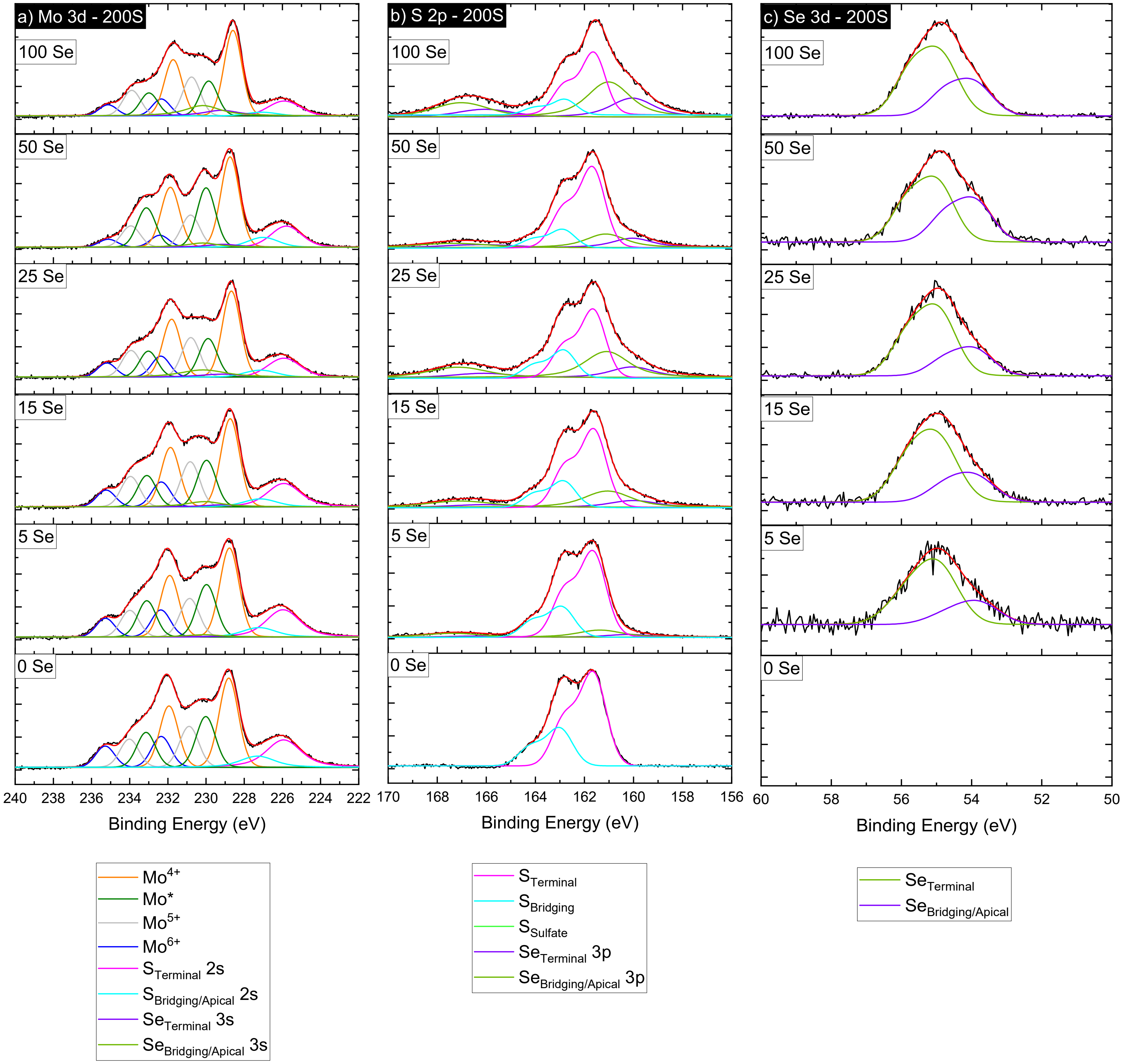

Figure S3: XPS of all samples within the 200S sample set, with a) Mo3d, b) S 2p, and c) Se 3d

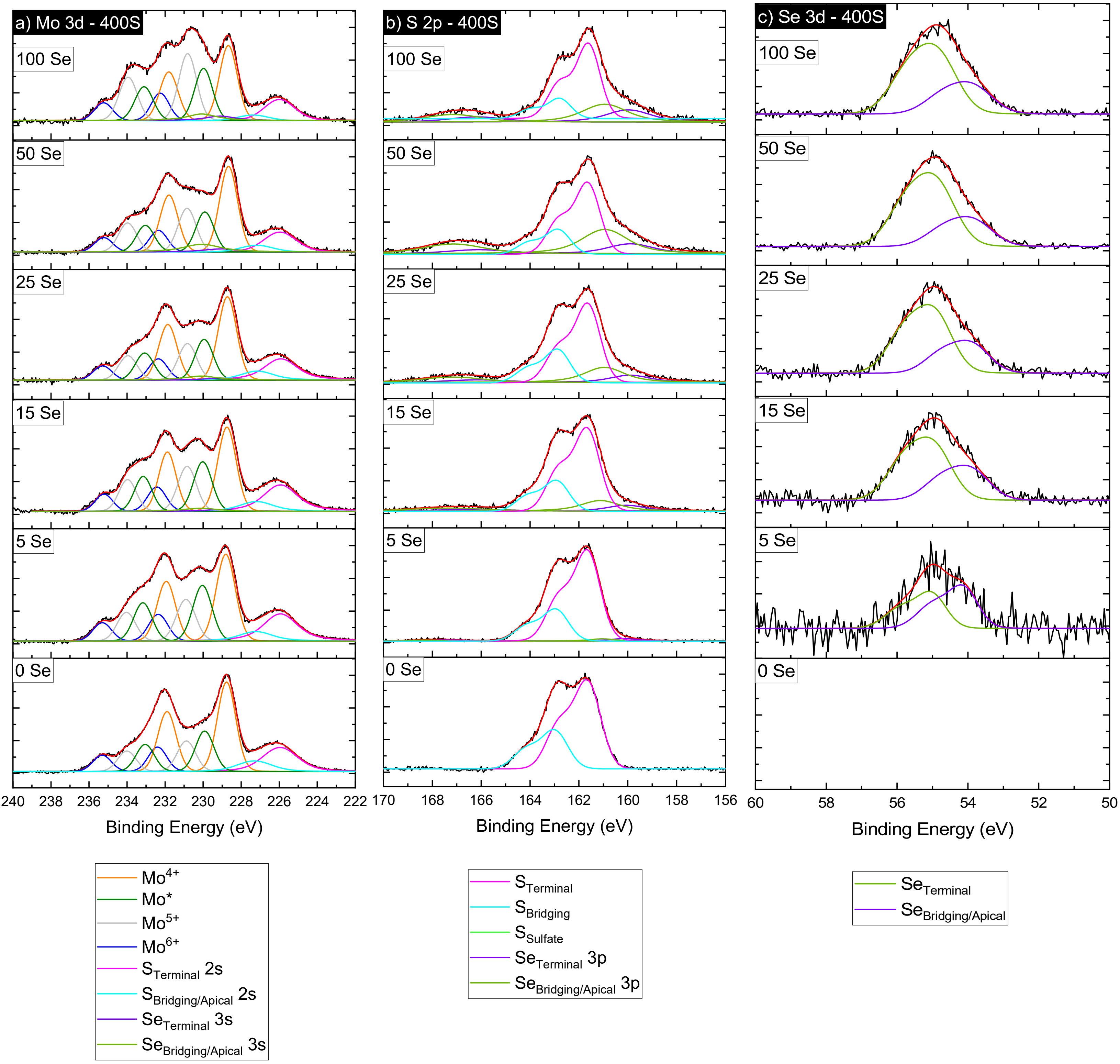

Figure S4: XPS of all samples within the 200S sample set, with a) Mo3d, b) S 2p, and c) Se 3d